\documentclass[twocolumn]{aastex63}
\usepackage{amssymb,amsmath}
\usepackage{graphicx}
\usepackage{xcolor,natbib}
\usepackage{multirow}
\usepackage[T1]{fontenc}
\usepackage{ae,aecompl}
\usepackage{newtxtext, newtxmath}
\usepackage{float}
\usepackage{subfigure}
\usepackage{longtable}
\usepackage{enumitem}
\usepackage{longtable,listings}
\usepackage[flushleft]{threeparttable}
\usepackage{parcolumns}
\def\oiiic{[O~{\sc iii}]$\lambda5007_c$\ }
\def\oiiib{[O~{\sc iii}]$\lambda5007_e$\ }
\def\ciii{[O~{\sc iii}]$\lambda4363_c$\ }
\def\biii{[O~{\sc iii}]$\lambda4363_e$\ }

\def\obj{SDSS J1058+5443}

\accepted{\today}
\submitjournal{ApJ}

\shorttitle{The best candidate of true type-2 quasar}
\shortauthors{Zhang X. G. \& Zhao S. D.}

\begin{document}

\title{SDSS J1058+5443: A blue quasar does not have optical/NUV broad emission lines}

%\correspondingauthor{XueGuang Zhang}%
%\email{xgzhang@njnu.edu.cn}
\author{XueGuang Zhang$^{*}$}
\affiliation{School of Physical Science and Technology, GuangXi University, No. 100,
        Daxue Road, Nanning, 530004, P. R. China}

\author{SiDan Zhao}
\affiliation{School of Physics and technology, Nanjing Normal University, No. 1,
        Wenyuan Road, Nanjing, 210023, P. R. China}

\begin{abstract} %%%about 194 words
	In the manuscript, the blue quasar SDSS J105816.19+544310.2 (=\obj) at redshift 0.479 is reported 
as so-far the best candidate of true type-2 quasar with disappearance of central BLRs. There are so-far 
no definite conclusions on the very existence of true type-2 AGN, mainly due to detected optical broad 
emission lines in high quality spectra of some previously classified candidates of true type-2 AGN. Here, 
not similar as previous reported candidates for true type-2 AGN among narrow emission-line galaxies with 
weak AGN activities but strong stellar lights, the definite blue quasar \obj~ can be well confirmed as a 
true type-2 quasar due to apparent quasar-shape blue continuum emissions but apparent loss of both the 
optical broad Balmer emission lines and the NUV broad Mg~{\sc ii} emission line. Based on different model 
functions and F-test statistical technique, after considering blue-shifted optical and UV Fe~{\sc ii} 
emissions, there are no apparent broad optical Balmer emission lines and/or broad NUV Mg~{\sc ii} line, 
and the confidence level is smaller than 1sigma to support broad optical and NUV emission lines. Moreover, 
assumed the Virialization assumption to broad line emission clouds, the re-constructed broad emission lines 
strongly indicate that the probable intrinsic broad emission lines, if there were, cannot be hidden or 
overwhelmed in the noises of SDSS spectrum of \obj. Therefore, \obj, so-far the best and robust candidate 
of true type-2 quasar, leads to the further clear conclusion on the very existence of true type-2 AGN. 
\end{abstract}


\section{Introduction}

	Type-1 AGN (broad line Active Galactic Nuclei) and type-2 AGN (narrow line AGN) are the two main 
classes of AGN, due to quite different spectroscopic emission line features. The constantly being improved 
Unified Model \citep{an93, nh15} has been widely accepted to explain the different observational phenomena 
between type-1 and type-2 AGN: central AGN continuum emissions and broad line emissions from central broad 
emission line regions (BLRs) are seriously obscured by central dust torus in type-2 AGN. And the well 
detected polarized broad emission lines in type-2 AGN, such as the results well discussed in \citet{am85, 
tr01, tr03, sg18}, provide robust evidence to strongly support the Unified Model, indicating that type-1 
and type-2 AGN have the same central geometric structures including similar BLRs, but central BLRs hidden 
in type-2 AGN.

	Interestingly, based on the pioneer work on studying polarized broad emission lines in type-2 AGN 
in \citet{am85, tr01, tr03, nk04, cm07} and then followed in \citet{sg10, bv14, zh14, ly15, pw16, zh21c}, 
there is one special kind of AGN: true type-2 AGN (or AGN with none hidden-BLRs, or BLRs-less AGN or 
unobscured type-2 AGN in the literature), which have no expected hidden central BLRs. There are about 30 
Seyfert 2 galaxies with no hidden BLRs reported in \citet{tr01, tr03}, due to lack of polarized broad 
emission lines. There are two strong candidates of true Type-2 AGN, NGC3147 and NGC4594, reported in 
\citet{sg10}, due to their few X-ray extinctions and the upper limits on the broad emission line luminosities 
are two orders of magnitude lower than the average of typical Type-1 AGN. A small number of Type-2 quasar 
are reported in \citet{bv14} through their strong g-band variabilities. A sample of candidates of true 
Type-2 AGN are reported in \citet{zh14}, through both their the long-term variability and the expected 
reliable power law components in their spectra in SDSS (Sloan Digital Sky Survey). A candidate of true 
Type-2 AGN, SDSS J0120, is reported in \citet{ly15}, through its long-term variability properties but 
none-detected broad emission lines. A small sample of candidates of true Type-2 AGN are reported in 
\citet{pw16} based on their properties of unobscured X-ray emissions. More recently, \citet{zh21c} reported 
the composite galaxy SDSS J1039 as a candidate of true type-2 AGN, based on its long-term variabilities 
and its apparent loss of broad optical Balmer emission lines.

	Meanwhile, there are different physical models proposed to explain the disappearance of central 
BLRs, such as the well reported models in \citet{nm13, eh09, cao10, ip15, en16}, strongly indicating 
disappearance of central normal BLRs depending on physical properties of central AGN activities and/or 
depending on properties of central dust obscuration. \citet{nm13} have shown that absence or presence 
of central BLRs can be well regulated by accretion rate, assumed the BLRs formed by accretion disk 
instabilities. \citet{eh09} have proposed a disk-wind scenario to predict the disappearance of the central 
BLRs at quite low luminosities. \citet{cao10} has shown that disappearance of central BLRs associated 
with the outflows from the accretion disks can be expected in AGN with Eddington ratio smaller than 0.001. 
\citet{en16} have shown that disappearance of central BLRs could be expected in AGN with higher 
luminosities than $4\times10^{46}{\rm erg~s^{-1}}$ considering mass conservations in the disk outflows.

	However, detailed study on reported candidates of true type-2 AGN questions the very existence 
of true type-2 AGN, such as the detailed study in Mrk573 and NGC3147. \citet {tr01} has classified 
Mrk573 as a candidate of true type-2 AGN, however \citet{nk04} have shown the clearly detected polarized 
broad Balmer lines, leading Mrk573 as a normal type-2 AGN with central hidden BLRs. Similarly, \citet{sg10, 
ba12} have classified NGC3147 as a candidate of true type-2 AGN through both spectropolarimetric results 
and unobscured X-ray emission properties, however \citet{ba19} more recently have clearly detected 
double-peaked broad H$\alpha$ in high quality HST spectrum. Furthermore, \citet{ip15} have well discussed 
that none detected polarized broad emission lines in candidates of some true type-2 AGN may probably 
due to effects of less scattering medium and/or due to effects of the increased torus obscurations.

	Besides true type-2 AGN with loss of central BLRs as a precious subclass of AGN, there is one 
another precious subclass of AGN: weal-line quasars which have quite weak broad emission lines in 
UV-optical band. Since the first sample of 74 high redshift ($z>3$) weak-line quasars in SDSS reported 
in \citet{df09}, weak-line quasars are interesting targets to provide further clues on accretion disk 
formation and/or abnormal properties of central BLRs for high ionization emission lines, such as in 
\citet{st10, ps15, ms18, pp22}. However, among the reported weak-line quasars identified by UV emission 
line properties, optical Balmer emission line (H$\alpha$ and H$\beta$) properties have been checked 
and reported in dozens of weak-line quasars, such as the results in \citet{ps15} and the reported virial 
BH masses of weak-line quasars by properties of broad Balmer lines in \citet{mn20}, indicating there are 
apparent broad Balmer emission lines (large rest-frame equivalent widths) in weak-line quasars. 
Therefore, weak-line quasars with quite weak UV emission lines (especially Ly$\alpha$ and C~{\sc IV}) 
but apparent broad Balmer lines are not similar as true type-2 AGN with none-detected broad Balmer 
lines, and there are no further descriptions and discussions on weak-line quasars in the manuscript.

	There is so-far no definite conclusion on the very existence of true type-2 AGN because of lower 
quality spectra with larger noises being applied to explain the undetected broad emission lines, but 
{\bf to detect true type-2 AGN in high luminous blue AGN} would provide more clearer clues on the very 
existence of true type-2 AGN. Here, we report one candidate of true type-2 AGN in the blue quasar \obj~ 
with none detected broad emission lines, which will provide robust evidence to support the very existence 
of true type-2 AGN. And due to none-detected broad Balmer lines, there are no further discussions on 
the \obj~ as one weak-line quasar in the manuscript. Section 2 presents our main results on spectroscopic 
properties of \obj, to provide strong evidence to support disappearance of broad optical and NUV emission 
lines. Section 3 gives properties of BH mass of \obj. Section 4 shows the necessary discussions. Section 5 
gives our final summaries and conclusions. And we have adopted the cosmological parameters 
$H_{0}=70{\rm km\cdot s}^{-1}{\rm Mpc}^{-1}$, $\Omega_{\Lambda}=0.7$ and $\Omega_{\rm m}=0.3$.

\begin{figure*}
\centering\includegraphics[width = 18cm,height=16cm]{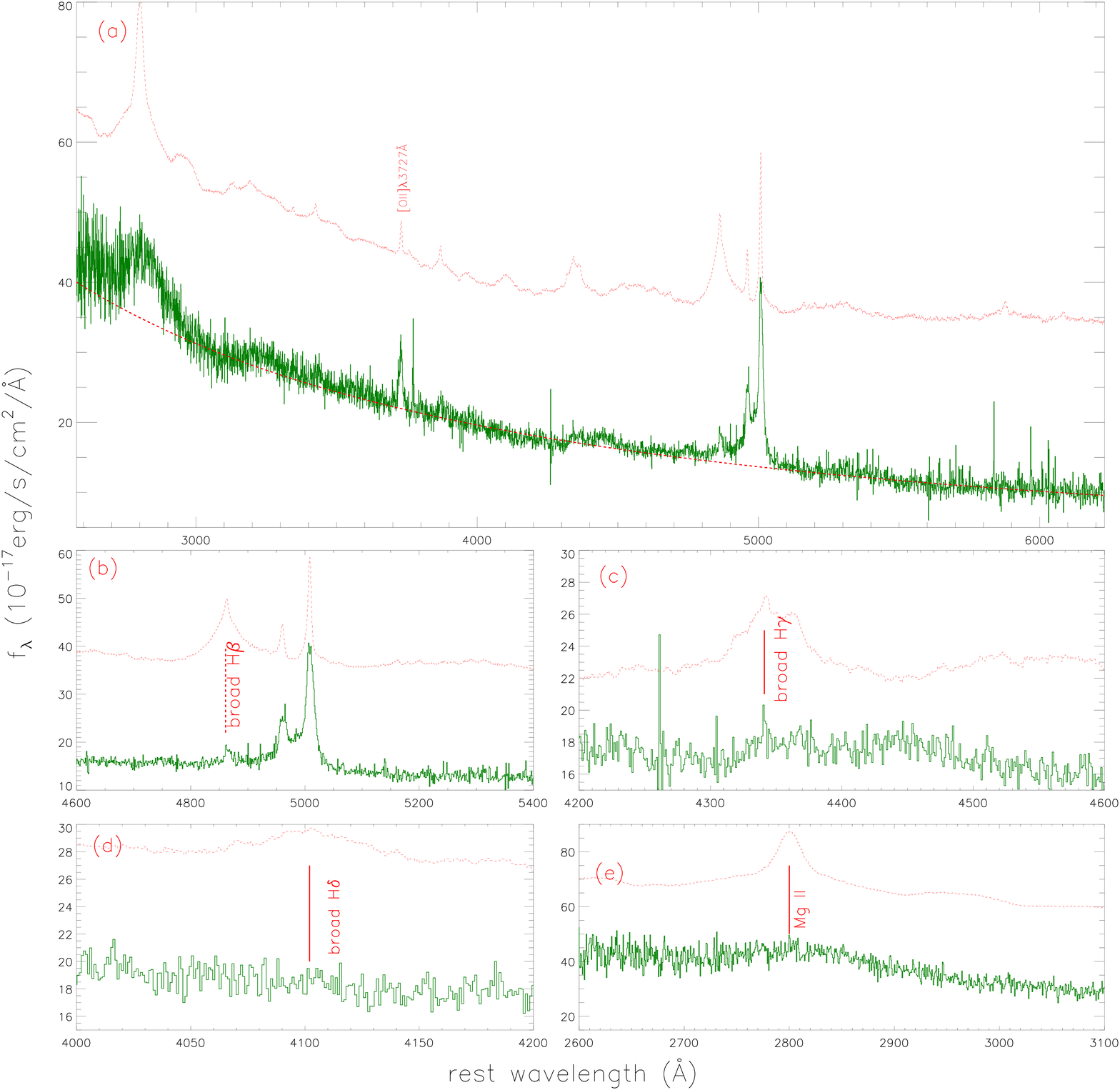}
\caption{{\bf Panel (a)} shows the whole SDSS spectrum of \obj~ (solid dark green line) and the composite 
spectrum of SDSS quasars (dashed pink line). Dashed red line shows the power law component to describe 
the continuum emissions. {\bf Panel (b), (c), (d) and (e)} show the spectrum (solid dark green line) 
around H$\beta$ (rest wavelength from 4600 to 5400\AA), H$\gamma$ (rest wavelength from 4250 to 4500\AA), 
H$\delta$ (rest wavelength from 4000 to 4200\AA) and Mg~{\sc ii} (rest wavelength from 2600 to 3100\AA) 
and the composite spectrum of SDSS quasars (dashed pink line), respectively. Vertical red lines mark the 
positions of the expected broad H$\beta$, broad H$\gamma$, broad H$\delta$ and broad Mg~{\sc ii} lines in 
the Panel (b), (c), (d) and (e).}
\label{spec}
\end{figure*}

\begin{figure*}
\centering\includegraphics[width = 18cm,height=6cm]{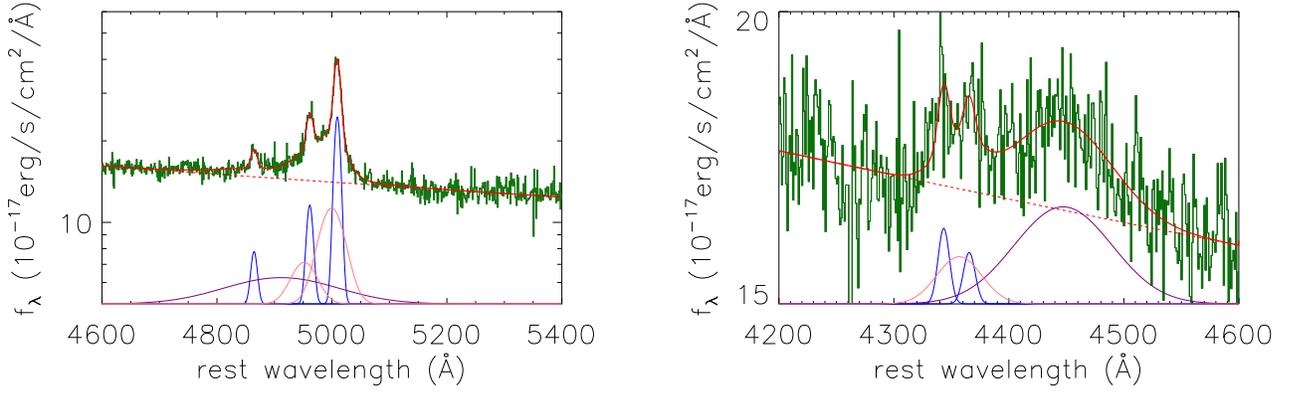}
\caption{Best-fitting results to the emission lines around H$\beta$ (left panel) and around H$\gamma$ (right 
panel) by model A. In each panel, solid dark green line shows the SDSS spectrum, solid red line shows the 
best-fitting results, dashed red line shows the determined power law continuum emissions, solid blue lines 
show the determined narrow Balmer line and core components of [O~{\sc iii}] emissions, solid pink lines show 
the determined extended components of [O~{\sc iii}] emissions, solid purple line shows the determined broad 
emission component. In order to show clearer results, the Y-axis is in logarithmic coordinate in each panel.
}
\label{hb}
\end{figure*}

\begin{figure*}
\centering\includegraphics[width = 18cm,height=10cm]{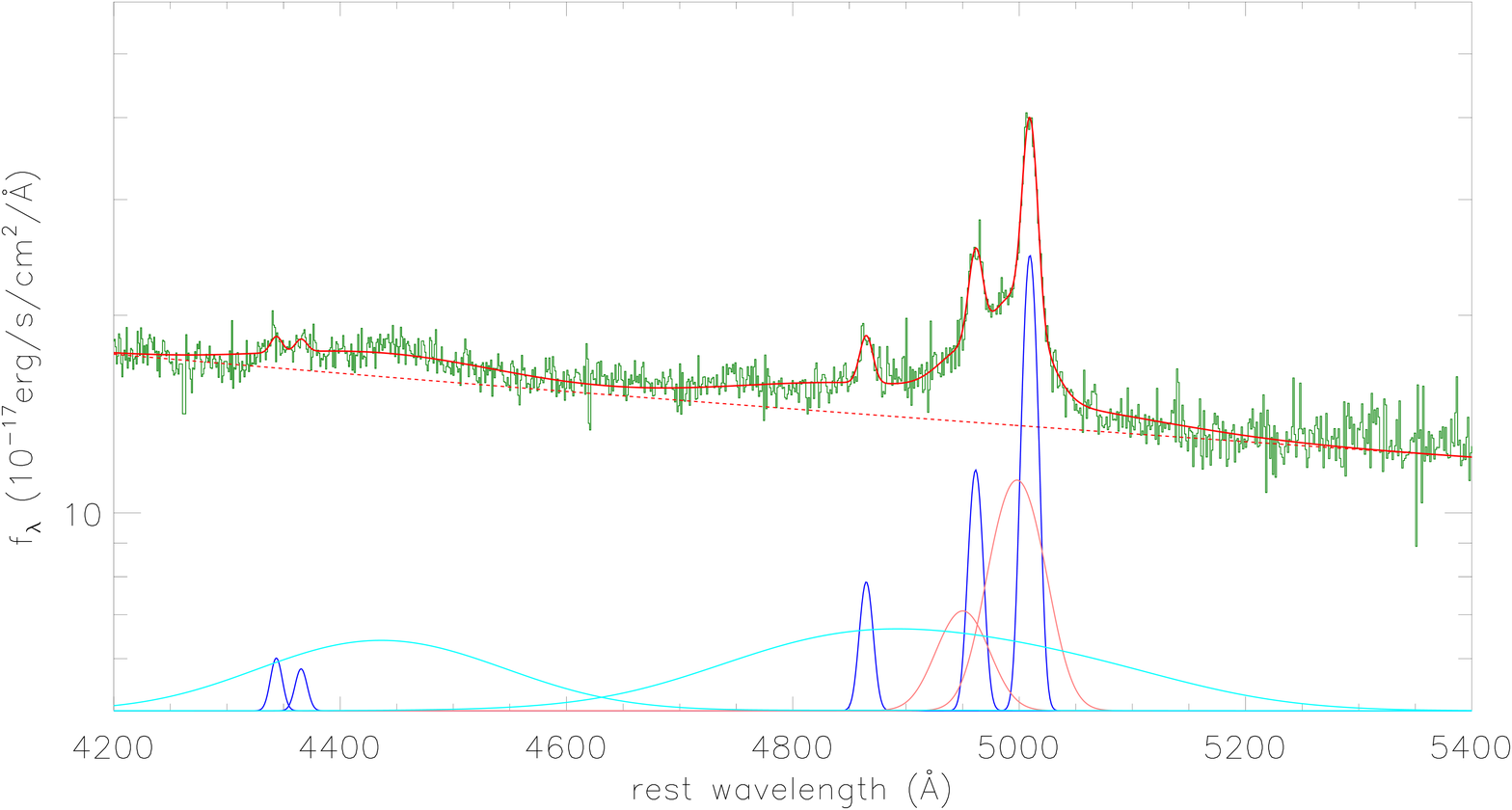}
\caption{Best-fitting results to the emission lines with rest wavelength from 4200 to 5400\AA~ by model B 
with considering optical Fe~{\sc ii} emissions. Solid dark green line shows the SDSS spectrum, solid red 
line shows the best-fitting results, dashed red line shows the determined power law continuum emissions, 
solid blue lines show the determined narrow Balmer line and core components of [O~{\sc iii}] emissions, solid 
pink lines show the determined extended components of [O~{\sc iii}] emissions, solid cyan lines 
show the determined optical Fe~{\sc ii} emissions related to the lower term of the transitions of $^4$F, 
$^6$S in \citet{kp10}. In order to show clearer results, the Y-axis is in logarithmic coordinate.
}
\label{feii}
\end{figure*}

\section{Spectroscopic properties of \obj}

%%%sec2

	\obj~ at redshift 0.479 is a well classified blue quasar in SDSS, with its spectrum shown in 
Fig.~\ref{spec}. The blue continuum emissions from NUV to NIR can be well described by a power law 
$f_\lambda\propto(\lambda)^{-1.65\pm0.01}$ with the continuum luminosity at rest wavelength 5100\AA~ to be 
(about $\lambda~L_{5100}\sim(5.74\pm0.06)\times10^{44}{\rm erg/s}$, leading \obj~ to be a definitely blue 
quasar. Meanwhile, the narrow emission lines, especially the apparent [O~{\sc iii}]$\lambda4959, 5007$\AA~ 
doublet, can be well applied to totally confirm the redshift $z\sim0.479$. Moreover, the composite spectrum 
of SDSS quasars reported in \citet{vb01} is also shown in Fig.~\ref{spec}, providing interesting clues on 
the apparent blue continuum emission features of \obj. More interestingly, besides the quasar-shape power 
law continuum emissions, it looks like that there are no broad optical Balmer emission lines and no broad 
NUV Mg~{\sc ii} line but apparent Fe~{\sc ii} emissions in the \obj, especially based on the shown detailed 
emission features around H$\beta$, around H$\gamma$, around H$\delta$ and around Mg~{\sc ii} line in the 
panel (b), (c), (d), and (e) in the Fig.~\ref{spec}.

\subsection{Properties of optical Balmer lines}

%%sec2.1
%%%%1
	Besides the direct comparisons between the SDSS spectroscopic emission line features and the emission 
line features in the composite spectrum of SDSS quasars, shown in Fig.~\ref{spec}, the detailed properties of 
the line emissions around the expected broad emissions lines are well measured by Gaussian functions as follows. 
Before proceeding further, there is one point we should note. There is a very broad emission component around 
4450\AA, however, so broad component is not existed around H$\beta$, indicating the broad component should be 
not from Balmer emission clouds. Therefore, in order to well check properties of the very broad component 
around 4450\AA, the emission properties with rest wavelength range from 4200 to 5400\AA~ are mainly checked, 
for the emission lines around H$\beta$ and H$\gamma$, by the following three different model functions.

%%%2
	For the model A which assumes the very broad component around 4450\AA~ coming from the Balmer emission 
regions, the following model functions are included. A narrow plus a broad Gaussian functions (second moment 
smaller or larger than 500${\rm km/s}$) $G_{H\beta,~N}([\lambda_{0,~\beta~N},~\sigma_{\beta~N},~f_{\beta~N}])$ 
(parameter $\lambda_{0}$ and $\sigma$ as central wavelength and second moment in units of \AA) and 
$G_{H\beta,~B}([\lambda_{0,~\beta~B},~\sigma_{\beta~B},~f_{\beta~B}])$ are applied to describe the H$\beta$. 
A narrow and a broad Gaussian functions $G_{H\gamma,~N}([\lambda_{0,~\gamma~N},~\sigma_{\gamma~N},~f_{\gamma~N}])$ 
and $G_{H\gamma,~B}([\lambda_{0,~\gamma~B},~\sigma_{\gamma~B},~f_{\gamma~B}])$ are applied to describe the 
H$\gamma$. Two narrow Gaussian functions $G_{c1}([\lambda_{0,~c1},~\sigma_{c1},~f_{c1}])$ and 
$G_{c2}([\lambda_{0,~c2},~\sigma_{c2},~f_{c2}])$ are applied to describe the core components of the 
[O~{\sc iii}]$\lambda4959,~5007$\AA~ doublet. Two broad Gaussian functions $G_{e1}([\lambda_{0,~e1},~\sigma_{e1},
~f_{e1}])$ and $G_{e2}([\lambda_{0,~e2},~\sigma_{e2},~f_{e2}])$ are applied to describe the extended components 
of the [O~{\sc iii}] doublet \citep{gh05a, sh11, zh21}, considering probably spatially extended 
emission regions for the outflow-related extended [O~{\sc iii}] components as discussed in \citet{za16}. A broad 
and a narrow Gaussian functions $G_{c3}([\lambda_{0,~c3},~\sigma_{c3},~f_{c3}])$ and 
$G_{e3}([\lambda_{0,~e3},~\sigma_{e3},~f_{e3}])$ are applied to describe the core and the extended components 
of the [O~{\sc iii}]$\lambda4363$\AA. A power law component $A\times(\frac{\lambda}{4100\textsc{\AA}})^B$ is 
applied to describe the continuum emissions underneath the emission lines. When the model functions are applied, 
the following restrictions are accepted. First, line flux and second moment of each Gaussian component are not 
smaller than zero. Second, ratios of the central wavelengths (in unit of \AA) and second moments (in unit of \AA) 
of $G_{H\beta,~N}$ to $G_{H\gamma,~N}$ are fixed to be 
\begin{equation}
%\begin{split}
\frac{\lambda_{0,~\gamma~N}}{4341.68\textsc{\AA}}~=~\frac{\lambda_{0,~\beta~N}}{4862.68\textsc{\AA}}
\ \ \ \ \ \ \ \ \frac{\sigma_{0,~\gamma~N}}{4341.68\textsc{\AA}}~=~\frac{\sigma_{0,~\beta~N}}{4862.68\textsc{\AA}}
%\end{split}
\end{equation}
leading the narrow Balmer lines to have the same redshift and the same line widths. Third, ratios of the central 
wavelengths, second moments and fluxes of $G_{c1}$ to $G_{c2}$ to $G_{c3}$ are fixed to be 
\begin{equation}
\begin{split}
&\frac{\lambda_{0,~c1}}{5008.24\textsc{\AA}}~=~\frac{\lambda_{0,~c2}}{4960.295\textsc{\AA}}
	~=~\frac{\lambda_{0,~c3}}{4364.436\textsc{\AA}}\\ 
&\frac{\sigma_{0,~c1}}{5008.24\textsc{\AA}}~=~\frac{\sigma_{0,~c2}}{4960.295\textsc{\AA}}
	~=~\frac{\sigma_{0,~c3}}{4364.436\textsc{\AA}} \\
&f_{c1}~=~3~\times~f_{c2}\ \ \ \ \  (f_{c3} \ \ free)
\end{split}
\end{equation}
Fourth, ratios of the central wavelengths, second moments and fluxes of $G_{e1}$ to $G_{e2}$ to $G_{e3}$ are fixed 
to be 
\begin{equation}
\begin{split}
&\frac{\lambda_{0,~e1}}{5008.24\textsc{\AA}}~=~\frac{\lambda_{0,~e2}}{4960.295\textsc{\AA}}
	~=~\frac{\lambda_{0,~e3}}{4364.436\textsc{\AA}} \\ 
&\frac{\sigma_{0,~e1}}{5008.24\textsc{\AA}}~=~\frac{\sigma_{0,~e2}}{4960.295\textsc{\AA}}
	~=~\frac{\sigma_{0,~e3}}{4364.436\textsc{\AA}} \\
&f_{e1}=3~\times~f_{e2}\ \ \ \ \  (f_{e3} \ \ free)
\end{split}
\end{equation}
The model parameters of $G_{H\gamma,~B}$ and $G_{H\beta,~B}$ are free model parameters, which will provide further 
clues to the origin of the very broad components around 4450\AA, from Balmer emissions or from extended [O~{\sc iii}] 
emissions or other physical origins. Then, through the Levenberg-Marquardt least-squares minimization technique, 
the best-fitting results to the emission lines by the model A can be well determined, and shown in Fig.~\ref{hb} 
with $\chi^2/dof\sim0.88$ (where $\chi^2$ and $dof$ are the summed squared residuals for the best-fitting results 
and the degree of freedom, respectively). Here, in order to show clearer results, left panel of Fig.~\ref{hb} shows 
the best-fitting results to the emission lines around H$\beta$ with rest wavelength range from 4600 to 5400\AA, and 
right panel of Fig.~\ref{hb} shows the best-fitting results to the emission lines around H$\gamma$ with rest 
wavelength range from 4200 to 4600\AA.

%%%%3

	Before proceeding further, one point is noted. Besides the two narrow Gaussian components for narrow H$\beta$ 
and narrow H$\gamma$ and two broad Gaussian components for broad H$\beta$ and broad H$\gamma$ discussed above, two 
Gaussian components are also considered in the Model A to describe probable 'outflow components' in narrow H$\beta$ 
and narrow H$\gamma$ (the extended component for wings of narrow Balmer lines, such as the broad Gaussian components 
included in [O~{\sc iii}] lines), and two additional Gaussian components are considered in Model A to describe probable 
'outflow components' in broad H$\beta$ and broad H$\gamma$ (or to describe probably complicated profiles of broad 
Balmer lines). However, for the model including the new four additional Gaussian components applied, the determined 
probable 'outflow components' in broad Balmer lines have fluxes to be zero, and the determined 'outflow components' 
in narrow Balmer lines have their line fluxes one times smaller than their corresponding uncertainties. Therefore, 
it is not necessary to consider the probable 'outflow components' in narrow Balmer line or in broad Balmer lines.

%%%4
	The determined parameters of emission lines are listed in Table~1 for the model A. The very broad component 
around 4450\AA~ can be well described by a broad Gaussian function with central wavelength 4447\AA, second moment 
42\AA, and there is an expected corresponding very broad component underneath of H$\beta$ described by a broad 
Gaussian function with central wavelength 4912\AA, second moment 95\AA. It is clear that the very broad components 
coming from very extended components of [O~{\sc iii}] emissions can be totally ruled out, due to the following main 
reason. If the very broad component around 4450\AA~ was from extended component of [O~{\sc iii}]$\lambda4363$\AA, 
the broad component is a red shifted component, however, for the very component around 4922\AA~ assumed from the 
extended component of [O~{\sc iii}]$\lambda4959, 5007$\AA~ doublet, it is a blue-shifted component. Therefore, 
the very broad component around 4450\AA~ should be probably from broad Balmer emission lines. However, there are 
quite different second moments and red-shifted velocities (relative to narrow Balmer lines) of the broad components 
of H$\beta$ and H$\gamma$, $\sigma_{\beta~B}\sim95$\AA~($\sim5860{\rm km/s}$) but 
$\sigma_{\gamma~B}\sim42$\AA~($\sim2900{\rm km/s}$), $\lambda_{0,~\beta_B}-\lambda_{0,~\beta_N}\sim47$\AA~ 
($\sim2900{\rm km/s}$) but $\lambda_{0,~\gamma_B}-\lambda_{0,~\gamma_N}\sim104$\AA~ ($\sim7170{\rm km/s}$), leading 
to incompatible dynamic properties of broad Balmer emission clouds. Moreover, the measured line width of the very 
broad component around 4450\AA~ can also be applied to rule out that the determined extended components of 
[O~{\sc iii}]$\lambda4959, 5007$\AA~ were actually from broad H$\beta$ emissions, because the extended component 
has line width about 23-25\AA, 1.9times smaller than the broad component around 4450\AA. Therefore, the very broad 
components around 4450\AA~ and around 4900\AA~ determined by the Model A are not preferred from broad Balmer emission 
clouds or from extended [O~{\sc iii}] emission clouds. Preferred choice should be mainly considered for the origin 
of the very broad components around 4450\AA~ and around 4910\AA.

%%%5
	For the model B, new model functions are considered as follows. Besides the narrow Balmer emission 
components and the components from the [O~{\sc iii}] emissions described the Gaussian functions ($G_{H\beta,~N}$, 
$G_{H\gamma,~N}$, $G_{c1}$, $G_{c2}$, $G_{c3}$, $G_{e1}$, $G_{e2}$, $G_{e3}$) and the AGN continuum emissions 
described by a power law function, there is an additional component from optical Fe~{\sc ii} emissions described 
by the broadened optical Fe~{\sc ii} template in \citet{kp10}, which are mainly considered to explain the very 
broad component around 4450\AA~ and around 4910\AA. Considering the optical Fe~{\sc ii} emission 
templates in four groups (three groups related to the lower term of the transitions of $^4$F, $^6$S and $^4$G 
terms, and the fourth group from the I Zw 1 Fe~{\sc ii} template) as discussed in \citet{kp10}, the Fe~{\sc ii} 
component $G_{Fe}$is described by 
\begin{equation}
G_{Fe}=\sum_{i=1}^{4}I_i~\times~T_{Fe,~i,~V_s,~V_b}
\end{equation}
where $I_i~\ge~0$ represents the intensity of Fe~{\sc ii} emission features in each group, $T_{Fe,~i}$ means the 
broadened and shifted Fe~{\sc ii} emission features in each group, and $V_s$ and $V_b$ means the same shift  
velocity and the same broadening velocity for the Fe~{\sc ii} emission features in the four groups. Then, the 
same restrictions applied in model A are accepted in model B on the Gaussian components of narrow Balmer emission 
components and the core and extended components from the [O~{\sc iii}] emissions. Then, through the 
Levenberg-Marquardt least-squares minimization technique, the best-fitting results to the emission lines by the 
model B can be well determined after considering the optical Fe~{\sc ii} emissions, and shown in Fig.~\ref{feii} 
with $\chi^2_B/dof_B\sim1330.88/1329$. The determined parameters of emission lines are also listed in Table~1 for 
the model B. The results by model B are more natural than those by model A, because there is no need to reconcile 
the contradiction on the origin of the very broad component around 4450\AA~ and around 4900\AA. Therefore, in 
the current stage, the model B is preferred in the \obj, indicating there are no broad Balmer emission lines 
in \obj.

\begin{table*}
\caption{Line parameters of the emission features}
\begin{tabular}{lccc|lccc}
\hline\hline
\multicolumn{4}{c}{Model A} & \multicolumn{4}{c}{Model B} \\
Line    &   $\lambda_0$  &   $\sigma$  & flux  & Line    &   $\lambda_0$  &   $\sigma$  & flux \\
\hline
H$\beta_N$  & 4864.56$\pm$0.65 & 5.45$\pm$0.68 & 38.29$\pm$4.86 &
H$\beta_N$  & 4864.91$\pm$0.66 & 5.79$\pm$0.61 & 41.23$\pm$4.42 \\
H$\beta_B$  & 4911.72$\pm$11.99 & 94.66$\pm$9.96 & 296.69$\pm$42.51 &
\dots & \dots & \dots & \dots \\
H$\gamma_N$ & 4343.36$\pm$0.58 & 4.87$\pm$0.61 & 14.16$\pm$4.96 &
H$\gamma_N$ & 4343.67$\pm$0.59 & 5.17$\pm$0.54 & 14.84$\pm$3.06 \\
H$\gamma_B$ & 4447.14$\pm$3.93 & 41.67$\pm$4.92 & 157.07$\pm$19.61 &
\dots & \dots & \dots & \dots \\
\oiiic & 5009.46$\pm$0.15 & 5.88$\pm$0.17 &  288.26$\pm$10.66 &
\oiiic & 5009.48$\pm$0.15 & 5.98$\pm$0.16 &  297.84$\pm$10.11 \\
\oiiib & 4999.51$\pm$1.13 & 22.25$\pm$1.18 & 350.95$\pm$19.03 &
\oiiib & 4996.34$\pm$1.17 & 22.69$\pm$0.96 & 357.56$\pm$13.78 \\
\ciii & 4365.49$\pm$0.13 & 5.12$\pm$0.15 &  10.02$\pm$5.22 &
\ciii & 4365.52$\pm$0.13 & 5.21$\pm$0.14 &  10.53$\pm$2.74 \\
\biii & 4356.82$\pm$0.98 & 19.39$\pm$1.03 & 34.65$\pm$14.44 &
\dots & \dots & \dots & \dots \\
\dots & \dots & \dots & \dots &
opt Fe~{\sc ii}$^*$ & -7702$\pm$256 & 6200$\pm$400 & 991$\pm$65 \\
\hline\hline
\end{tabular}\\
Notice: \oiiic~ and \oiiib~ represent the core and the extended component in the [O~{\sc iii}]$\lambda5007$\AA. 
\ciii~ and \biii~ represent the core and the extended component in the [O~{\sc iii}]$\lambda4363$\AA. The 
second and sixth columns show the determined rest central wavelengths in unit of \AA~ ($km/s$ for the shifted 
velocities of the optical Fe~{\sc ii}), the third and seventh columns show the determined line widths (second 
moment) in unit of \AA~ (the broadening velocity in unit of $km/s$ for the optical Fe~{\sc ii} lines), and 
the fourth and the eighth columns list the determined line flux in the unit of $10^{-17}{\rm erg/s/cm^2}$. 
The flux of optical Fe~{\sc ii} is total flux of the blue-shifted optical Fe~{\sc ii} emissions from 4200\AA~ 
to 5400\AA~ in \obj.
\end{table*}

%%%%6

	Furthermore, as what we have known that the theoretical flux ratio of H$\beta$ to H$\gamma$ is about 2. 
The determined flux ratios of narrow H$\beta$ to narrow H$\gamma$ are about $2.70_{-0.96}^{+1.98}$ and 
$2.78_{-0.72}^{+1.09}$ by model A and by model B, respectively, consistent with the theoretical value, providing 
further clues to support the reliability of the determined narrow H$\beta$ and narrow H$\gamma$ components by the 
model A and by the model B.

%%%%7
	Moreover, as the shown and determined Fe~{\sc ii} emission features by model B in Table~1 and in 
Fig.~\ref{feii}, Fe~{\sc ii} emission features in the two groups related to the lower term of the transitions 
of $^4$F, $^6$S in \citet{kp10} are strongly apparent, but there are quite weak Fe~{\sc ii} emission features 
related to the groups related to the lower term of the transition of $^4$G in \citet{kp10} and the Fe~{\sc ii} 
template from I Zw 1. Different intensity ratios of optical Fe~{\sc ii} features in the four groups in \citet{kp10} 
can be found in the shown examples in the webpage \url{http://147.91.240.26/FeII_AGN/link6.html} maintained by 
Dr. Kovacevic. And the determined optical Fe~{\sc ii} emission features have blue-shifted velocity about 
$-7700{\rm km/s}$ and have EW (Equivalent Width) about 64\AA. The blue-shifted velocity is so-far the largest 
blue-shift velocity among the AGN with optical Fe~{\sc ii} lines. However, as discussed in \citet{kp10} (see 
their Fig.~17), there is a negative correlation between EW of optical Fe~{\sc ii} and EW of 
[O~{\sc iii}]$\lambda5007\AA$. In \obj, the EW 48\AA~ of [O~{\sc iii}]$\lambda5007\AA$ and the EW 64\AA~ of 
optical Fe~{\sc ii} are roughly consistent with the EW correlation between optical Fe~{\sc ii} and 
[O~{\sc iii}]$\lambda5007\AA$, to provide further clues to support the determined optical Fe~{\sc ii} emission 
features. Furthermore, as discussed above, only Fe~{\sc ii} features in two of the four groups in \citet{kp10} 
are strong enough, therefore, the other Fe~{\sc ii} templates, such as discussed in \citet{bg92, vj04, tk06}, 
can not efficiently applied to describe the optical Features in \obj, due to partly fixed intensity ratios of 
optical Fe~{\sc ii} features in their templates..

%%%%8
	For the model C, based on the results determined by model B, there are two additional broad Gaussian 
components, $G_{H\beta}([\lambda_{0,~\beta},~\sigma_{\beta},~f_{\beta}])$ and $G_{H\gamma}([\lambda_{0,~\gamma},
~\sigma_{\gamma},~f_{\gamma}])$, in model C applied to describe probable broad Balmer lines, besides the 
optical Fe~{\sc ii} emissions. When, the model C is running, the additional restrictions are accepted on 
central wavelengths and second moments of $G_{H\beta}$ and $G_{H\gamma}$, 
\begin{equation}
%\begin{split}
	\frac{\lambda_{0,~\gamma}}{4341.68\textsc{\AA}}~=~\frac{\lambda_{0,~\beta}}{4862.68\textsc{\AA}}
\ \ \ \ \ \frac{\sigma_{0,~\gamma}}{4341.68\textsc{\AA}}~=~\frac{\sigma_{0,~\beta}}{4862.68\textsc{\AA}}
%\end{split}
\end{equation}
to confirm that the probable broad Balmer lines have the same redshift and the same line widths. Meanwhile, 
when the model C is running, the determined model parameters of the broadening and shift velocities of 
optical Fe~{\sc ii} template are accepted as the starting values in model C, and to assure that the optical 
Fe~{\sc ii} emissions are larger than zero, otherwise, the final results come to be the totally same results 
as those by model A. Then, through the Levenberg-Marquardt least-squares minimization technique, the 
best-fitting results to the emission lines by the model C can be well determined after considering the 
probable broad Balmer lines, with $\chi^2_C/dof_C\sim1326.69/1325$. The determine broad H$\beta$ can be 
describe by a Gaussian component with central wavelength $4890.68\pm45.23$\AA, second moment $32.69\pm41.56$\AA, 
and flux $13.54\pm21.83$. And the determined broad H$\gamma$ with line flux zero. The best-fitting results 
are not shown in plots, but are totally similar as the results shown in Fig.~\ref{feii}, besides the only 
quite weak broad Gaussian component around 4890.68\AA. Based on the determined line flux and line width 
quite smaller than their corresponding uncertainties of $G_{H\beta}$, the determined broad $G_{H\beta}$ is 
not reliable enough.

%%%9
	Moreover, the F-test technique can be applied to determine that the broad Gaussian component 
$G_{H\beta}$ determined by model C is not necessary to be considered. Based on the different $\chi^2/Dof$ 
values for the model C and Model B for optical spectroscopic emissions with and without considerations 
of probable broad Balmer emission lines, the calculated $F_p$ value is about
\begin{equation}
F_p=\frac{\frac{\chi^2_C-\chi^2_B}{dof_C-dof_B}}{\chi^2_C/dof_C}\sim1.046
\end{equation}
Based on $dof_C-dof_B$ and $dof_C$ as number of dofs of the F distribution numerator and denominator, 
the expected value from the statistical F-test with confidence level about 62\% will be near to $F_p$. 
Therefore, the confidence level is only about 62\%, smaller than 1sigma, to support the broad H$\beta$  
component, indicating the determined broad Gaussian component by model C is not preferred.

%%%%10
	Therefore, based on the shown and discussed results above, no optical broad emission lines can 
be preferred in \obj:
\begin{itemize}
\item If there were broad H$\beta$ and H$\gamma$ determined by model A, the expected broad H$\beta$ 
and broad H$\gamma$ have significantly different both red-shifted velocities and line widths, leading 
to incompatible dynamic properties of broad Balmer emission clouds. 
\item The determined extended components of [O~{\sc iii}] doublet are truly from [O~{\sc iii}] 
emissions, not part of broad H$\beta$ emissions. 
\item Considering blue-shifted optical Fe~{\sc ii} emissions, optical spectroscopic emission features 
can be naturally described.
\item The F-test technique can be well applied to confirm that it is not necessary to consider broad 
Balmer lines, once the optical Fe~{\sc ii} emissions are fully considered.
\end{itemize}

\subsection{Properties of NUV Mg~{\sc ii} line}

%%%1
	Besides properties of the optical Balmer emission lines, emission lines around Mg~{\sc ii} with 
rest wavelength range from 2600 to 3600\AA~ can be well measured by two different models as follows. Here, 
the selected wavelength range extending to 3600\AA~ can lead to well determined continuum emissions. 

\begin{figure*}
\centering\includegraphics[width = 18cm,height=6cm]{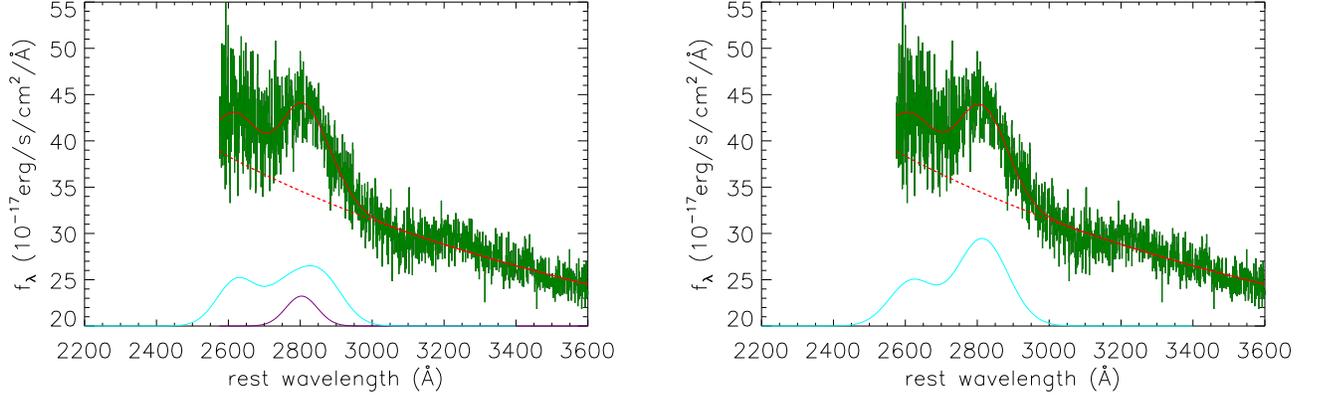}
\caption{Best fitting results to the emissions around 2800\AA~ by Model A (left panel) and model B (right panel). 
In each panel, solid dark green line shows the SDSS spectrum, solid red line shows the best-fitting results, 
dashed red line shows the determined power law continuum emissions, solid cyan line shows the determined UV 
Fe~{\sc ii} emissions. In left panel, solid purple line shows the determined broad Gaussian component of 
Mg~{\sc ii} by model A.
}
\label{mg2}
\end{figure*}

%%%2
	In model A, a broadened and shifted UV Fe~{\sc ii} template discussed in \citet{kp15} plus a 
broad Gaussian function and a power law component are applied to describe the emissions. Here, 
considering the UV Fe~{\sc ii} templates in four groups (4 multiplets: 60 within 2907 to 2979\AA, 61 within 
2861 to 2917\AA, and 62 and 63 which overlap within 2709 to 2749\AA) as discussed in \citet{kp15}, the 
UV Fe~{\sc ii} component $G_{uFe}$is described by
\begin{equation}
	G_{uFe}=\sum_{i=1}^{4}I_i~\times~T_{uFe,~i,~V_{us},~V_{ub}}
\end{equation}
where $I_i~\ge~0$ represents the intensity of UV Fe~{\sc ii} emission features in each group, $T_{uFe,~i}$
means the broadened and shifted UV Fe~{\sc ii} emission features in each group, and $V_{us}$ and $V_{ub}$ 
means the same shifted velocity and the same broadening velocity for the UV Fe~{\sc ii} emission features 
in the four groups. The best-fitting results are shown in the left panel of Fig~\ref{mg2} with 
$\chi^2_A/dof_A\sim1345.73/1442\sim0.93$ by model A. The determined broad Mg~{\sc ii} has the central 
wavelength $2803.61\pm13.48$\AA, second moment $43.86\pm25.82$\AA~ ($\sim4700{\rm km/s}$), and line flux 
$(355.01\pm531.71)\times10^{-17}{\rm erg/s/cm^2}$. Considering the measured line flux smaller than the 
corresponding uncertainty, the determined broad Mg~{\sc ii} component is not reliable enough. The measured 
broadening velocity and blue-shifted velocity of the UV Fe~{\sc ii} emissions are about $5888\pm965{\rm km/s}$ 
and $-12820\pm1376{\rm km/s}$. And, the UV Fe~{\sc ii} emission features in the two groups 
from multiplets 60 and 62 are more strongly apparent than the features from the multiplets 61 and 63.

%%%3
	In model B with no considerations of broad Mg~{\sc ii} component, only the broadened and shifted 
UV Fe~{\sc ii} template discussed in \citet{kp15} plus a power law component are applied to describe the 
emissions lines. The best-fitting results are shown in right panel of Fig~\ref{mg2} with 
$\chi^2_B/dof_B\sim1347.47/1445\sim0.93$ by model B. The measured broadening velocity and blue-shifted   
velocity of the UV Fe~{\sc ii} emissions are about $6757\pm239{\rm km/s}$ and $-14140\pm1155{\rm km/s}$. 
And, the UV Fe~{\sc ii} emission features in the two groups from multiplets 60 and 62 are 
more strongly apparent than the features from the multiplets 61 and 63. Meanwhile, if the UV Fe~{\sc ii} 
components including apparent effects of central radial flows, the larger blue-shifted velocity in UV 
Fe~{\sc ii} emissions than in optical Fe~{\sc ii} emissions can be well accepted, after accepted UV 
Fe~{\sc ii} emissions are from regions nearer to central power source. Model B not considering a broad 
Gaussian component can also lead to well accepted best-fitting results to the emission lines around 2800\AA.  

%%%4
	Not only the large uncertainties of line flux and second moments as clues to rule out the expected 
broad Mg~{\sc ii} component, the F-test technique can be also applied to determine that the broad Gaussian 
component is not necessary to be considered. Based on the different $\chi^2/Dof$ values for the model A and 
Model B for emission lines around 2800\AA, the calculated $F_p$ value is about
\begin{equation}
F_p=\frac{\frac{\chi^2_A-\chi^2_B}{dof_A-dof_B}}{\chi^2_A/dof_A}\sim0.62
\end{equation}
Based on $dof_A-dof_B$ and $dof_A$ as number of dofs of the F distribution numerator and denominator, the 
expected value from the statistical F-test with confidence level about 40\% will be near to $F_p$. Therefore, 
the confidence level is only about 40\%, smaller than 1sigma, to support the broad Mg~{\sc ii} component, 
indicating the determined broad Gaussian component by model A is not preferred. Certainly, the discussed 
results above are only based on different model functions and corresponding F-test technique, further efforts 
are necessary to totally confirm the discussed results above.

%%%5
	Furthermore, as discussed in \citet{kp15} (see left panel in their Fig.~10), there is one positive 
broadening velocity correlation between optical Fe~{\sc ii} emission features and UV Fe~{\sc ii} features. 
Based on the determined broadening velocity about 6700${\rm km/s}$ of UV Fe~{\sc ii} emission features and 
the determined broadening velocity about 6200${\rm km/s}$ of optical Fe~{\sc ii} emission features in \obj, 
there are clues to consider \obj~ to simply follow the broadening velocity correlation between UV and 
optical Fe~{\sc ii} emission features.

\section{BH mass of \obj}

	BH mass, one of fundamental parameters of AGN, is mainly estimated in the subsection. BH mass 
properties will be used in the following subsections to provide further clues to confirm probable loss 
of central BLRs in \obj. Due to no apparent broad emission lines in \obj, the reported virial BH mass 
is not considered in the manuscript based on the Virialization assumption applied to broad line emission 
clouds \citep{pf04, sh11}. The following two methods are mainly considered.

	First, based on the dependence of BH masses on continuum luminosity reported in Equation (9) in 
\citet{pf04}, BH mass of \obj~ can be estimated as 
\begin{equation}
M_{BH}\propto(\lambda L_{5100})^{0.79\pm0.09}\sim3.02_{-0.84}^{+1.17}\times10^8{\rm M_\odot} 
\end{equation}	
with BH mass uncertainty determined by uncertainty of continuum luminosity and uncertainties 
of the factors in the equation.

\begin{figure}
\centering\includegraphics[width = 8cm,height=5.5cm]{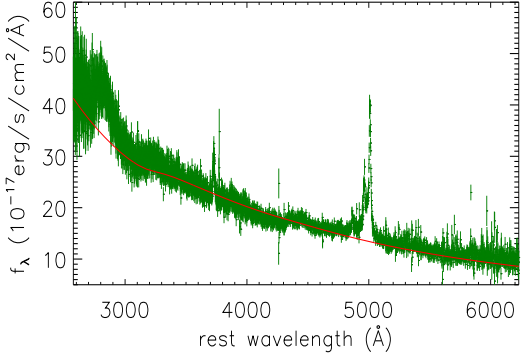}
\caption{The best descriptions to the SDSS continuum emissions through the well applied AGNSPEC code. Dots 
plus error bars in dark green show the SDSS spectrum, solid red line shows the determined best descriptions 
to continuum emissions.}
\label{asp}
\end{figure}

	Second, the commonly accepted AGNSPEC code \citep{ha00, dw07} is applied to roughly determine the 
central BH mass through properties of spectral energy distributions of \obj, considering the non-LTE general 
relativistic accretion disk models. Fig.~\ref{asp} shows the best descriptions to the SDSS spectrum of 
\obj~ with emission line features around 2800\AA~ (from 2700\AA~ to 2950\AA) and around 4900\AA~ (from 
4800\AA~ to 5100\AA) being masked out, with the BH mass $M_{BH}\sim(2.5\pm1.1)\times10^8{\rm M_\odot}$, 
accretion rate $\dot{M}\sim0.03\pm0.01{\rm M_\odot/year}$, BH spin $a_*\sim0.91\pm0.06$, and inclination angle 
$\cos(i)\sim0.32\pm0.11$, through the AGNSPEC code created templates applied through the 
Levenberg-Marquardt least-squares minimization technique.

	The two different methods lead to similar BH mass, therefore the mean value 
$M_{BH}\sim(2.8\pm1.4)\times10^8{\rm M_\odot}$ is accepted as the BH mass of \obj~ in the manuscript\footnote{
The virial BH mass about $3\times10^9{\rm M_\odot}$ is reported in \citet{sh11} in \obj. However, as discussed 
above, the viral BH mass sensitively depending on broad line properties should be not reliable enough in \obj. 
Therefore, rather than the virial BH mass $3\times10^9{\rm M_\odot}$, the mass $M_{BH}\sim2.8\times10^8{\rm M_\odot}$ 
is accepted in \obj~ in the manuscript. Moreover, after checked the results in \citet{sh11}, there is weak 
correlation between virial BH mass and narrow line width in SDSS quasars, with Spearman rank correlation 
coefficient only about 0.07. Therefore, in the manuscript, we do not consider line width (second moment) 
$350\pm15{\rm km/s}$ of core component of [O~{\sc iii}]$\lambda5007$\AA~ to be accepted as substitute of 
stellar velocity dispersion, and then to estimate BH mass by the M-sigma relation \citep{fm00, ge00, kh13, 
mm13, bt15, sg15, bb17}.}.

\section{Main Discussions}

\subsection{Re-constructed probable intrinsic but hidden broad Balmer lines?}

	Based on the estimated BH mass, the intrinsic broad Balmer lines could be re-constructed. Then, it 
is interesting to check whether the expected intrinsic broad Balmer lines are actually overwhelmed in 
noises of SDSS spectrum of \obj.

	Not similar as the previous reported candidates of true type-2 AGN, weak broad emission components 
could lead to none detected broad emission lines in low quality spectra with large noises. In the blue quasar 
\obj, strong power law AGN continuum emissions clearly indicate strong intrinsic broad Balmer emission lines, 
if there were broad line emissions. Based on the strong positive linear correlation (scatter of 
0.2dex) between AGN continuum luminosity and total Balmer line luminosity from SDSS quasars in \citet{gh05}, 
the expected total H$\beta$ luminosity (including both the broad and the narrow components in H$\beta$) of 
\obj~ could be 
\begin{equation}
	L(H\beta)\propto(\lambda L_{5100})^{1.133}\sim(1.02\pm0.48)\times10^{43}{\rm erg/s}
\end{equation}
with uncertainty $0.48\times10^{43}{\rm erg/s}$ determined by the accepted scatter of 0.2dex of 
the correlation above. Considering the measured line luminosity of narrow H$\beta$ about 
$(3.57\pm0.38)\times10^{41}{\rm erg/s}$ of \obj, the expected broad H$\beta$ luminosity (the 
total H$\beta$ luminosity minus the narrow H$\beta$ luminosity) should be about 
$(9.84\pm4.82)\times10^{42}{\rm erg/s}$, the corresponding broad H$\beta$ line flux is 
$(1132\pm550)\times10^{-17}{\rm erg/s/cm^2}$, about 27 times stronger than the narrow H$\beta$ 
in \obj. And due to the strong blue quasar-shape continuum emissions in \obj, there are no obscurations 
considered on the line flux of emission lines.

\begin{figure}
\centering\includegraphics[width = 8cm,height=5.5cm]{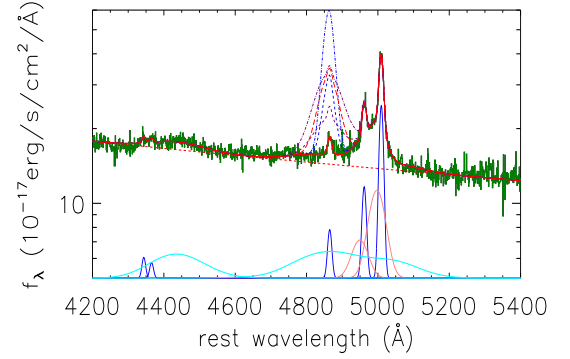}
\caption{The expected intrinsic broad H$\beta$ through the properties of virial BH mass in \obj. Symbols and 
line styles have the same meanings as those shown in Fig.~\ref{feii}, but the dashed red line, dot-dashed 
purple line, dashed purple line, dot-dashed blue line and dashed blue line show the re-constructed H$\beta$ after 
considering the intrinsic broad H$\beta$ emission features (if there were) with Gaussian parameters 
[$\lambda_0/$\AA,~$\sigma_B(H\beta)/({\rm km/s})$,~$f_B(H\beta)/(10^{-17}{\rm erg/s/cm^2})$] as 
[4862, 1700, 1132], [4862, 1700+700, 1132+550], [4862, 1700+700, 1132-550], [4862, 1700-700, 1132+550], 
[4862, 1700-700, 1132-550], respectively.} 
\label{hb2}
\end{figure}

	If the expected broad H$\beta$ with line luminosity of $(9.84\pm0.24)\times10^{42}{\rm erg/s}$  
was intrinsically true in \obj, there was only one point leading the expected strong broad H$\beta$ not to be 
detected in the SDSS spectrum: the strong broad H$\beta$ was extremely broad enough that the expected broad H$\beta$ 
with very lower heights are totally overwhelmed by the noises of SDSS spectrum. However, considering the estimated 
BH mass $M_{BH}\sim(2.8\pm1.4)\times10^8{\rm M_\odot}$, line width $\sigma_B$ of the expected intrinsic 
broad Balmer lines can be well estimated through the Virialization assumption \citep{pf04} combing with size-luminosity 
empirical relation \citep{bd13} (scatter of 0.13dex) to estimate distance $R_{BLRs}$ of BLRs to central BH, 
\begin{equation}
\begin{split}
R_{BLRs}~&\propto~L_{5100}^{0.542}\sim93\pm28{\rm light-days} \\
\sigma_B(H\beta)~&=~(\frac{G~M_{BH}}{5.5~R_{BLRs}})^{0.5}\sim1700\pm700{\rm km/s}
\end{split}
\end{equation}
where the factor $5.5$ is the applied scale factor to estimate Virial BH mass used in \citet{pf04}, 
and uncertainty 28{\rm light-days} of $R_{BLRs}$ is determined by the scatter of 0.13dex of the 
size-luminosity empirical relation, and uncertainty ${\rm 700km/s}$ of $\sigma_B(H\beta)$ is determined by the 
uncertainties of both BH mass and $R_{BLRs}$. Based on second moment $\sigma_B(H\beta)=1700\pm700{\rm km/s}$ 
and line flux $f_B(H\beta)=(1132\pm550)\times10^{-17}{\rm erg/s/cm^2}$, the expected intrinsic Gaussian-like broad 
H$\beta$ $G_{H\beta,~N}([\lambda_0, \sigma_B(H\beta), f_B(H\beta)])$ with central wavelength $\lambda_0=4862$\AA~ 
can be well re-constructed and shown in Fig.~\ref{hb2}, after considering uncertainties of parameters 
of $\sigma_B(H\beta)$ and $f_B(H\beta)$. It is clear that the expected intrinsic broad H$\beta$ is strong enough, 
if there was, the intrinsic broad H$\beta$ can not be overwhelmed by noises of SDSS spectrum.

\subsection{Re-constructed probable intrinsic but hidden broad Mg~{\sc ii} line?}

	Similar as the procedure to re-construct intrinsic broad H$\beta$ (if there was), the intrinsic broad 
Mg~{\sc ii} can also be re-constructed, to check whether the intrinsic broad Mg~{\sc ii} line can be hidden 
or overwhelmed in the noises of SDSS spectrum, if there was intrinsic broad Mg~{\sc ii} line.

	First, based on the strong positive linear correlation (scatter of 0.15dex) between AGN 
continuum luminosity at 3000\AA~ and Mg~{\sc ii} line luminosity from SDSS quasars in \citet{sh11} and the measured 
continuum luminosity at 3000\AA~ about $L_{3000}\sim(8.95\pm0.55)\times10^{44}{\rm erg/s}$ in \obj, the expected 
Mg~{\sc ii} line luminosity of \obj~ could be
\begin{equation}
L(Mg)\propto(\lambda L_{3000})^{0.909}\sim(1.11\pm0.43)\times10^{43}{\rm erg/s}
\end{equation}
with uncertainty $0.43\times10^{43}{\rm erg/s}$ determined by the accepted scatter of 0.15dex 
of the correlation above, leading the corresponding broad Mg~{\sc ii} line flux to be 
$(1280\pm490)\times10^{-17}{\rm erg/s/cm^2}$.

\begin{figure}
\centering\includegraphics[width = 8cm,height=5.5cm]{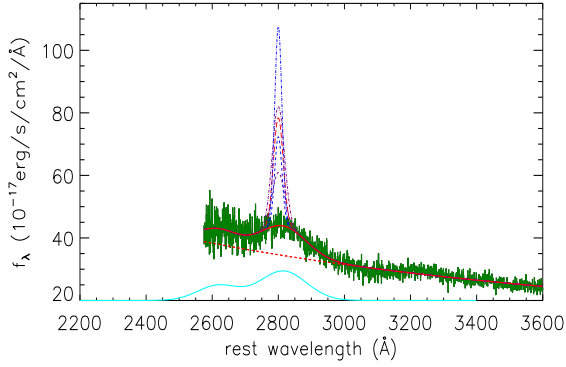}
\caption{The expected intrinsic broad Mg~{\sc ii} through the properties of virial BH mass in \obj. Symbols 
and line styles have the same meanings as those shown in right panel of Fig.~\ref{mg2}, but the dashed 
red line, dot-dashed purple line, dashed purple line, dot-dashed blue line and dashed blue line show the re-constructed 
intrinsic broad Mg~{\sc ii} emission features (if there were) after considering the intrinsic broad Mg~{\sc ii} 
emission features (if there were) with Gaussian parameters [$\lambda_0/$\AA,~$\sigma(Mg)/({\rm km/s})$,
~$f(Mg)/(10^{-17}{\rm erg/s/cm^2})$] as [2800, 1590, 1280], [2800, 1590+400, 1280+490], [2800, 1590+400, 1280-490], 
[2800, 1590-400, 1280+490], [2800, 1590-400, 1280-490], respectively.
}
\label{mg3}
\end{figure}

	Second, accepted the Virialization assumption to broad Mg~{\sc ii} emission clouds as discussed 
in \citet{sh11}, the line width (FWHM, full width at half maximum) can be estimated as 
\begin{equation}
FWHM(Mg)~=~(\frac{M_{BH}}{5.5~(\frac{L_{3000}}{\rm 10^{44}erg/s})^{0.62}})^{0.5}\sim3750\pm950{\rm km/s}
\end{equation}
with uncertainty $950{\rm km/s}$ of $FWHM(Mg)$ determined by the uncertainties of both BH mass and 
$L_{3000}$, leading to second moment about $1590\pm400{\rm km/s}$ assuming a Gaussian broad Mg~{\sc ii} line. 
Then, based on second moment $\sigma(Mg)=1590\pm400{\rm km/s}$ and line flux $f(Mg)=(1280\pm490)\times10^{-17}{\rm erg/s/cm^2}$ 
and assumed central wavelength 2800\AA, the expected intrinsic Gaussian-like broad Mg~{\sc ii} can be well 
re-constructed and shown in Fig.~\ref{mg3}, after considering uncertainties of parameters
of $\sigma(Mg)$ and $f(Mg)$. It is clear that the expected intrinsic broad Mg~{\sc ii} is strong enough, if there 
was, the intrinsic broad Mg~{\sc ii} cannot be overwhelmed by noises of SDSS spectrum.

\begin{figure*}
\centering\includegraphics[width = 18cm,height=10cm]{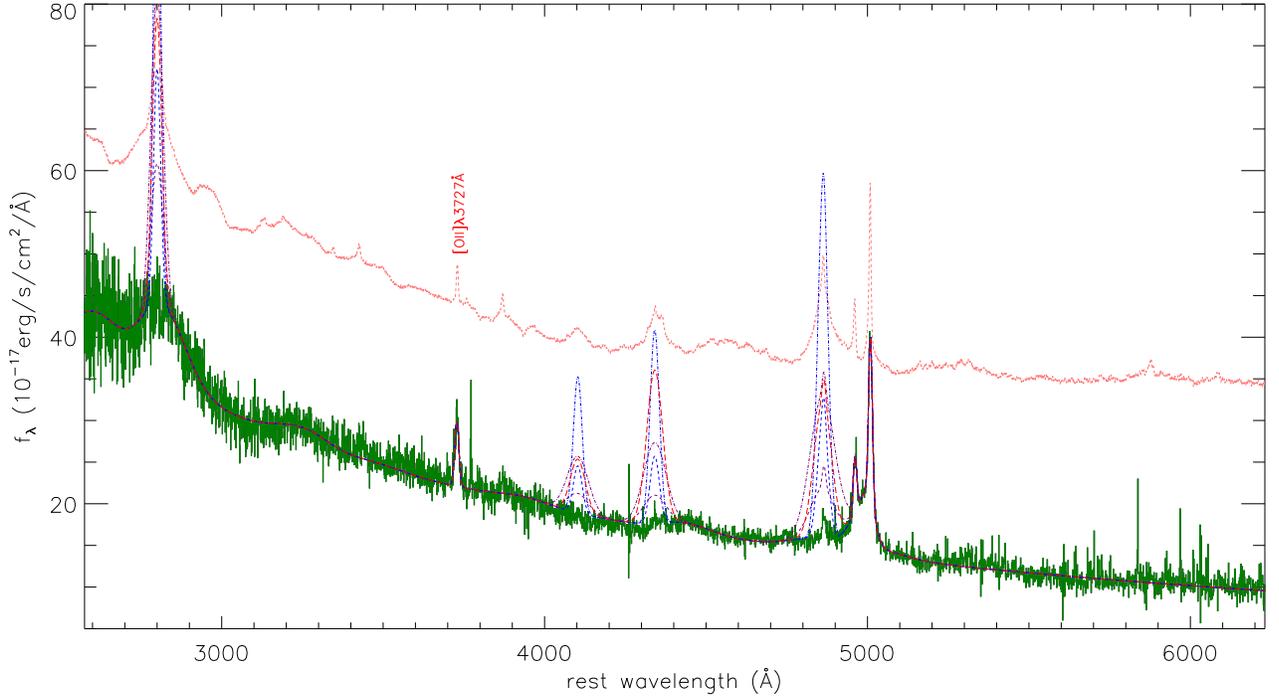}
\caption{The comparison between composite spectrum of SDSS quasars and the spectrum of \obj~ considering the 
probable intrinsic broad lines (if there were) under the Virialization assumption applied to broad line clouds. 
Dashed pink line shows the composite spectrum of SDSS quasars, solid dark green line shows the SDSS spectrum 
of \obj. Solid red lines around 5000\AA~ and around 3727\AA~ show the best descriptions to the 
[O~{\sc iii}]$\lambda4959, 5007$\AA~ and [O~{\sc ii}]$\lambda3727$\AA~ emission lines. Around 4861\AA, 4341\AA, 4103\AA~ 
and 2800\AA, dashed red lines, dot-dashed purple lines, dashed purple lines, dot-dashed blue lines and dashed blue 
line show the re-constructed intrinsic broad lines (if there were), after considering uncertainties of line widths 
and line fluxes, similar as shown in Fig.~\ref{hb2} and in Fig.~\ref{mg3}.
}
\label{comp}
\end{figure*}

	Now, based on the re-constructed broad Balmer lines and Mg~{\sc ii} line, if there were intrinsic broad 
emission lines, it is interesting to compare the expected SDSS spectrum of \obj~ considering the probable 
intrinsic broad lines and the composite spectrum of SDSS quasars. The comparison is shown in Fig.~\ref{comp}. 
Here, the expected intrinsic broad H$\gamma$ and H$\delta$ are re-constructed based on the re-constructed broad 
H$\beta$ with line flux ratios to be $f_{4862}:f_{4341}:f_{4103}=1:0.5:0.3$ ($f_{4862}=1132\times10^{-17}{\rm erg/s/cm^2}$) 
and with the same line width (1700${\rm km/s}$) in velocity space and with the same redshift. It is clear that the 
spectroscopic features with considering the expected intrinsic broad lines are totally similar as those of composite 
spectrum of SDSS quasars. In one word, if there were intrinsic broad lines, the broad lines cannot be hidden or 
overwhelmed in noises of SDSS spectrum in \obj. Therefore, it can be clearly confirmed the loss of broad lines in the \obj.

\subsection{Further discussions}

	Besides the detailed discussions on emission lines, origin of blue continuum emissions is simply 
considered in the subsection, in order to confirm that \obj~ is truly an AGN.

\begin{figure}[htp]
\centering\includegraphics[width = 8cm,height=4.6cm]{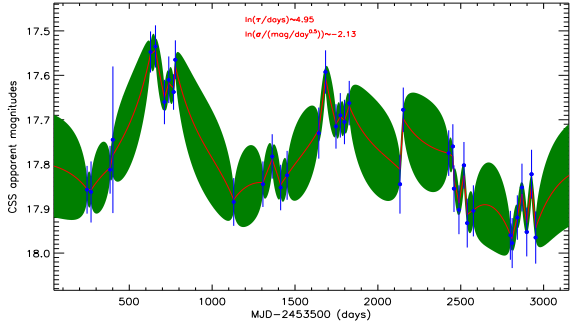}
\caption{Long-term variabilities of \obj~ from the CSS. Solid circles plus error bars in blue show the 
collected data points from the CSS, solid red line and the area covered by light green show the best 
descriptions and the corresponding 1sigma confidence bands to the light curve by the JAVELIN code. The 
determined characteristic variability timescale and amplitude are marked in the top corner.
}
\label{drw}
\end{figure}

	There are two further points applied to confirm that the blue continuum emissions are not from 
younger stellar objects but from the central AGN activities in \obj. On the one hand, long-term variabilities 
of \obj~ is well checked through the collected 8years-long light curve from the CSS (the Catalina Sky Survey) 
\citep{dd09} shown in Fig.~\ref{drw}. The apparent variabilities can be well described by the Damped Random 
Walk (DRW) process \citep{kb09, koz10, zk13} which has been proved to be the preferred process to describe 
intrinsic AGN variabilities. Through the public code of JAVELIN, the determined DRW process parameter of the 
characteristic variability timescale is about $\ln(\tau/days)\sim4.95\pm0.54$ ($\tau\sim140days$), similar 
as the mean value of $\tau$ in SDSS quasars well discussed in \citet{mi10}. On the other hand, the determined 
flux ratio of the total [O~{\sc iii}]$\lambda5007$\AA~ (both the core and broad component) to the narrow 
H$\beta$ is about 16.3 larger than 10, indicating strong central AGN activities in \obj~ by applications of 
BPT diagrams \citep{kn19, zh20}. Therefore, the apparently blue continuum emissions are confirmed to be from 
central AGN activities. And then, due to the strong blue continuum emissions from central AGN activities, 
there are no discussions on the loss of broad emission lines due to \obj~ classified as a changing-look AGN 
at dim state \citep{to76, la15, yw18, zh21b}.

	Moreover, the none detected broad emission lines are not due to \obj~ as a BL Lac \citep{pa08}. The 
radio properties of \obj~ is well checked through the Faint Images of the Radio Sky at Twenty-cm (FIRST) 
\citep{hw15}. The radio intensity at 1.4GHz is about 17.7mJy in \obj, leading \obj~ to be a radio loud AGN 
with radio loudness about 29. However, the [O~{\sc iii}]$\lambda5007$\AA~ has rest equivalent width (EW) 
about 49\AA~ very larger than 5\AA, indicating \obj~ is not a BL Lac object but a normal radio loud quasar.

\subsection{What should be done in the near future?}

	Actually, there is one surefire way to confirm the \obj~ as a true type-2 AGN. To get the spectroscopic 
observations around H$\alpha$ will provide robust evidence to support the conclusions in the manuscript for 
\obj. If there were no broad components of H$\alpha$, there will be no doubt to well identify the \obj~ as a 
true type-2 AGN. We wish there will be good chance to get NIR spectroscopic observations of the \obj in the 
near future.

\subsection{Physical origin of the loss of central normal BLRs in \obj}

	Once loss of broad emission lines can be confirmed, it is interesting to consider the potential 
physical origin of the loss of central normal BLRs. In \obj~ with bolometric luminosity about 
$6-9\times10^{45}{\rm erg/s}$ (about 10 times of the continuum luminosity $\lambda L_{5100}$) \citep{nh19}, 
both the steep bluer power law AGN continuum emissions and the strong bolometric luminosity indicate that 
the disk-wind scenario with luminosities higher than $10^{46}{\rm erg/s}$ as well discussed in \citet{en16} 
can be well applied to explain the loss of broad emission lines in \obj, rather than the accreting process 
with lower accretion rates as discussed in \citet{cao10, nm13, eh09}, etc. And the apparent blue-shifted 
broad UV and OPT Fe~{\sc ii} emission lines and the AGNSPEC simply determined large $a_*\sim0.91$ can provide 
probable clues to support the expected disk-winds scenario in \citet{en16}.

	Before the end of the subsection, one point should be noted. As discussed in \citet{vp20}, 
there is a special subclass of AGN, the called core-extremely red quasars (core-ERQs) previously defined in 
\citet{hz17}, representing an intermediate evolutionary phase in which a heavily obscured quasar is blowing 
out the circumnuclear interstellar medium, leading to commonly consequent extreme outflows in high luminosity 
core-ERQs. In other words, the conditions of the disk-wind scenario discussed in \citet{en16} could be well 
satisfied in high luminosity core-ERQs, however, there are common BLRs in core-ERQs. Therefore, the proposed 
disk-wind scenario discussed in \citet{en16} can be applied to explain the disappearance of central BLRs, 
but cannot be accepted as a standard criterion leading to disappearance of central BLRs in AGN.

\section{Summaries and Conclusions}
Finally, the main conclusions are given as follows.
\begin{itemize}
\item \obj~ is a blue quasar with apparently blue continuum emissions coming from central AGN activities, based 
	on DRW process well determined long-term variabilities and flux ratio larger than 10 of [O~{\sc iii}] to 
	narrow H$\beta$ of \obj. Not similar as previous reported candidates of true type-2 AGN among emission-line 
	objects with weak AGN activities but strong host galaxy contribution, properties of \obj~ could provide 
	more robust evidence to support the very existence of true type-2 AGN.
\item  Considering different model functions applied to describe emission lines around H$\beta$, rather than broad 
	Balmer lines but blue-shifted optical Fe~{\sc ii} emissions can be preferred in \obj, leading to the 
	conclusion that there are no broad optical Balmer emission lines in \obj. The confidence level is smaller 
	than 1sigma to support probably existence of broad Balmer emission lines.
\item Considering different model functions applied to describe emission lines around 2800\AA~ in \obj, the 
	blue-shifted UV Fe~{\sc ii} emissions can be preferred in \obj, leading to the conclusion that there are 
	no broad NUV Mg~{\sc ii} line in \obj. The confidence level is smaller than 1sigma to support probably 
	existence of broad NUV Mg~{\sc ii} line.
\item  Considering Virialization assumption to broad emission clouds in \obj, combining with estimated BH mass, 
	the expected intrinsic broad optical Balmer lines and NUV Mg~{\sc ii} line can be well re-constructed, 
	if there were intrinsic broad emission lines in \obj. The SDSS spectrum plus re-constructed broad lines 
	are totally similar as the composite spectrum of SDSS quasars.
\item  The re-constructed intrinsic broad lines can not be hidden or overwhelmed in the noises of SDSS spectrum 
	of \obj, indicating the loss of broad emission lines are not due to spectral quality.
\item  The \obj~ is so far the best candidate of true type-2 quasar, leading to the clear answer to the very 
	existence of true type-2 AGN. And the expected disk-winds scenario with high luminosity could be preferred 
	to explain the loss of central BLRs in \obj.
\end{itemize}

\section*{Acknowledgements}
Zhang and Zhao gratefully acknowledge the anonymous referee for giving us constructive comments and 
suggestions to greatly improve our paper. Zhang gratefully acknowledges the financial support of NSFC-12173020. 
The manuscript has made use of the data from the SDSS (\url{https://www.sdss.org/}) funded by the Alfred P. 
Sloan Foundation, the Participating Institutions, the National Science Foundation and the U.S. Department of 
Energy Office of Science. The manuscript has made use of the long-term variability data from the CSS 
(\url{http://nesssi.cacr.caltech.edu/DataRelease/}). The manuscript has made use of the NASA/IPAC Extragalactic 
Database (NED) (\url{http://ned.ipac.caltech.edu/classic/}). The manuscript has made use of the VLA FIRST 
Survey (http://sundog.stsci.edu/). The manuscript has made use of the public JAVELIN code to describe long-term 
intrinsic variability properties of AGN (\url{http://astro.sjtu.edu.cn/~yingzu/codes.html}). The manuscript 
has made use of the AGNSPEC code to describe SEDs from accretion disks around central black holes 
(\url{https://github.com/jhmatthews/agnspec_grids}). 

\iffalse
\section*{Data Availability}
The data underlying this article will be shared on reasonable request to the corresponding 
author (\href{mailto:xgzhang@njnu.edu.cn}{xgzhang@njnu.edu.cn}).
\fi

%\label{lastpage}
\end{document}